\documentclass[aps,prd,10pt,reprint,nofootinbib,floatfix]{revtex4-1}

\usepackage[dvipsnames]{xcolor}
\usepackage{xparse}
\usepackage{graphicx}
\usepackage{amsmath,amssymb,amsfonts,amsbsy}
\usepackage{dcolumn}
\usepackage{rotating}
\usepackage{multirow}
\usepackage{makecell}
\usepackage{comment}
\usepackage{overpic}
\usepackage{lipsum}
\usepackage{verbatim}
\usepackage{cancel}
\usepackage{makecell}
\usepackage{enumitem}
\usepackage[english]{babel}
\usepackage[utf8]{inputenc}
\usepackage[normalem]{ulem} % for \sout, doesn't change \emph
\usepackage{silence}
\usepackage{url}
\usepackage[colorlinks=true, allcolors=Blue]{hyperref}
\usepackage{orcidlink}

\usepackage{soul}

\makeatletter
\newcommand*{\rom}[1]{\expandafter\@slowromancap\romannumeral#1@}
\makeatother

\allowdisplaybreaks%

\newif\ifshowdraft%
\showdrafttrue%      % final version: hide draft notes, show "final only"

\NewDocumentEnvironment{finaltext}{+b}{%
  \ifshowdraft\else
    {\begingroup\color{black}#1\endgroup}%
  \fi
}{}

\NewDocumentEnvironment{drafttext}{+b}{%
  \ifshowdraft%
    {\begingroup\small\color{blue}\begin{quote}\textbf{[Draft note:]}~#1\end{quote}\endgroup}%
  \fi
}{}

\begin{document}

\title{A Spectrum of Cosmological Rips and Their Observational Signatures}

\author{Mikel Artola\orcidlink{0009-0007-9068-1995}}
\email[Contact address: ]{mikel.artola@ehu.eus}
\author{Ruth Lazkoz\orcidlink{0000-0001-5536-3130}
}
\email[Contact address: ]{ruth.lazkoz@ehu.eus}
\affiliation{Department of Physics, University of the Basque Country UPV/EHU, P.O. Box 644, 48080 Bilbao, Spain}
\affiliation{EHU Quantum Center, University of the Basque Country UPV/EHU, 48940 Leioa, Spain}
\author{Vincenzo Salzano\orcidlink{0000-0002-4905-1541}}
\email[Contact address: ]{vincenzo.salzano@usz.edu.pl}
\affiliation{Institute of Physics, University of Szczecin, Wielkopolska 15, 70--451 Szczecin, Poland}
\date{\today}

\begin{abstract}
We present a unified dark energy framework capable of generating a continuous spectrum of cosmological “rip” scenarios---including the Big Rip, Grand Rip, Mild Rip, Little Rip, Little Sibling of the Big Rip, and the newly found Dollhouse Rip---while ensuring a physically consistent evolution across cosmic history. Building on earlier phenomenological proposals, we introduce a barotropic equation-of-state parameter with a sigmoid-like correction to guarantee a strictly positive dark energy density and to avoid early-time pathologies commonly present in previous models. Using this formulation, closed-form analytic expressions for the energy density can be obtained. This, in turn, enables a systematic classification of future singularities based on the signs and magnitudes of two key parameters of the model. We test these scenarios with state-of-the-art cosmological probes, including DESI DR2 BAO, cosmic chronometers, CMB compressed likelihoods, and the Pantheon+ supernovae sample. According to our Bayesian analysis, all rip scenarios yield best-fit parameters compatible with $\Lambda$CDM at the $1\sigma$ level, with Bayes factors weakly favoring $\Lambda$CDM. The mild, logarithmic evolution of the proposed dark energy density prevents current observations from distinguishing among the different future fates. We conclude that, for rip cosmologies to gain observational support over $\Lambda$CDM, they must display more accentuated late-time dynamical features---such as perhaps rapid transitions or a phantom-divide crossing---within the redshift range probed by present surveys.
\end{abstract}

\maketitle

\section{Introduction}

Cosmology has been concerned with the ultimate fate of the Universe for as long as it exists. In the current view, the dynamics at cosmological scales is dictated by the energy content subject to the laws of General Relativity (GR) or its extensions, and the elusive component known as dark energy (DE)~\cite{Huterer:1998qv} plays a key role in the destiny of the Universe. The existence of this component of the cosmic budget --- commonly accepted as the driver of the current accelerated expansion of the Universe --- was first revealed by Type Ia supernovae (SN) observations~\cite{SupernovaSearchTeam:1998fmf, SupernovaCosmologyProject:1998vns}. Stronger evidence supporting DE later emerged from increasingly precise constraints on other cosmological parameters, driven by measurements of the cosmic microwave background (CMB)~\cite{Boomerang:2000wrj, WMAP:2003gmp} and baryon acoustic oscillations (BAO)~\cite{SDSS:2005xqv, 2dFGRS:2005yhx}. This new paradigm of scientific uncertainty brought upon the possibility of radically new dynamical features in the asymptotically late universe.

The longest-standing and hardest-to-beat version of DE is a cosmological constant ($\Lambda$). In this case, the necessarily negative pressure $p$ preserves a constant ratio with respect to the energy density $\rho$, and the Universe reaches an asymptotic de~Sitter phase. In this framework, once acceleration kicks in --- slightly after the onset of DE dominance ---, it lasts forever but in a contained fashion: there is no destruction of cosmic structures and no future catastrophes arise. That said, widening the scope to allow for evolutionary equation-of-state parameters $w = p/\rho$ (unlike the cosmological constant, for which $w_\Lambda = -1$) opens the door to more dramatic endings.

The latter require a phantom behavior, $w < -1$, all the way to perhaps just the very final moment of the history of the Universe. Therefore, violations of the null and weak energy conditions~\cite{Hawking:1973uf} will occur and deep implications follow. In particular, in phantom models the energy density of this negative pressure component grows with time at all states; as a result, one of several possible abrupt ends of the Universe could take place. These dramatic events are typically addressed as cosmological ``doomsdays'' because of the dissociation of (some) structures they entail.

The pioneering representative of such frameworks, which has driven much of the interest in the study of a plethora of cosmological singularities, is also the most extreme: the Big Rip~\cite{McInnes:2001zw, Caldwell:2003vq}. It is characterized by a simultaneous blow up of the scale factor $a$, the Hubble rate $H$, and its time derivative $\dot{H}$ at a finite cosmic time. The consequence is that hierarchically all structures (from the biggest to smallest) will be torn apart~\cite{Nesseris:2004uj}. This means that galaxy clusters, galaxies, planetary systems, and ultimately atoms, will stay no longer in a ``compact'' equilibrium. This is therefore a sudden, catastrophic and universal rip.

Ensuing works explored less striking possibilities in which the outcome is postponed (and softened). The Little Rip~\cite{Frampton:2011sp} describes a universe in which the expansion rate grows without bound but the disintegration of structures occurs only asymptotically as $a$, $H$ and $\dot{H}$ blow up as $t \to \infty$. Unlike in the Big Rip, there is no finite-time singularity in the Little Rip, but this is no obstacle for the ultimate destruction of all bound systems.

A further refinement is the Little Sibling of the Big Rip~\cite{Bouhmadi-Lopez:2014cca}. This case preserves the divergence of $a$ and $H$ at infinite time, combined with a finite $\dot{H}$ at all cosmic times. The disruption is therefore softer than in the Little Rip, yet still leads to the eventual dissolution of all structures.

While the former cases never cease to be phantom-like, another case which is commonly framed alongside the previous one is worth mentioning for completeness. This is the Pseudo Rip~\cite{Frampton:2011aa}, characterized by an equation-of-state parameter $w$ that stays below $-1$ but approaches the $w = -1$ boundary asymptotically and sufficiently rapidly. This final de~Sitter state is crucial since the Hubble factor gets saturated and its derivative becomes null. As a consequence, some structures undergo severe disruption, but the most tightly held systems endure. Therefore, even though this constitutes yet another baleful event, it is somewhat alleviated because the associated rip is, so to say, ``filtered''.

In addition to the latter standard scenarios, a plethora of other singularities have been proposed withing cosmological frameworks. Their classification --- based on their strength to exert tidal forces on extended bodies --- has been addressed attending to various criteria~\cite{Tipler:1977zza, Krolak:1986pno}. Broadly speaking, singularities can be divided into strong and weak: strong singularities are those which can distort finite objects, whereas weak ones usually involve divergences only in certain physical quantities. Well-known strong singularities include the Big Bang and Big Rip, as well as the Grand Rip and Grand Crunch~\cite{Fernandez-Jambrina:2014sga} --- which resemble  their Big Rip/Bang counterparts except that $w \to -1$ as the singularity is approached ---, and the so-called Directional Singularities~\cite{Fernandez-Jambrina:2007ohv} --- which are accessible only to causal geodesics except for those with zero linear momentum. On the other hand, examples of weak singularities include Sudden singularities~\cite{Barrow:2004xh}, where the DE pressure diverges at finite time while the energy density remains finite; Big Freeze singularities~\cite{Bouhmadi-Lopez:2004mpi, Sen:2005sk, Bouhmadi-Lopez:2006fwq}, where both quantities diverge at finite time; Generalized Sudden singularities~\cite{Nojiri:2005sx}, in which both the energy density and pressure of DE vanish but higher derivatives of the Hubble factor blow up; and $w$-singularities~\cite{Dabrowski:2009kg, Fernandez-Jambrina:2010pep, Dabrowski:2014fha}, characterized by a divergence solely in the equation-of-state parameter. A recent review of advances in the study of cosmological singularities, together with further references to those discussed here and beyond, can be found in~\cite{Trivedi:2023zlf}.

This taxonomy illustrates that the cosmic endpoints implied by evolving DE raise theoretical difficulties. Relatedly, realizations of the scenarios presented in the literature often suffer from early-time pathologies, such as negative energy densities and unwanted divergences~\cite{Bouhmadi-Lopez:2014cca, Stefancic:2004kb}, or instabilities in the perturbation sector~\cite{Barrow:2009df}. In addition, the strong phantom behavior inferred often lacks a clear connection to an underlying equation-of-state parameter, and usually stems from oversimplified parameterizations. This suggests that the physical models may benefit from additional theoretical robustness.

Moving to observational cosmology, recent results from the Dark Energy Spectroscopic Instrument (DESI) collaboration~\cite{DESI:2024mwx, DESI:2025zgx} have strengthened the interest on this topic given the new dominant paradigm of evolving DE.\@ Those observations appear to prefer phantom-like behavior at early times, while also hinting at a recent crossing of the phantom divide~\cite{Ozulker:2025ehg, Scherer:2025esj, Gonzalez-Fuentes:2025lei}. Within the widely used Chevalier-Polarski-Linder (CPL) parameterization~\cite{Chevallier:2000qy, Linder:2002et}, this preference is realized in an intriguing way: CPL often yields a phantom behavior in the past that loses its phantom nature as the Universe expands. However, is this evolution an artifact of the chosen parameterization or does it manifest in more imaginative choices? Whether this is a genuine physical trend or an artifact of CPL's linear form remains an open question. 

It is therefore natural to explore new classes of models that pursue the essential phantom phenomenology aforementioned, consistent with the most up-to-date observational hints. Our main goal is to ensure a consistent early-time evolution of DE, with strictly positive energy density and stability, and at the same time provide a unified picture of a (possible) cascade of future cosmological rips. 

The remainder of this work is structured as follows. In Section~\ref{sec:the_model} we establish the cosmological framework and present the proposed family of uniparametric DE equations of state. We provide an analytical solution to its energy density, which remains positive over all cosmological history, and discuss the shortcomings of previous proposals regarding  observational cosmology and similar constructions to our main proposal. Section~\ref{sec:asymptotic_behavior_of_dark_energy} is devoted to the study of early- and late-time asymptotic behavior of DE depending on  the signs of the parameters introduced. Here we shall discuss the parameter-space region for which our model could yield catastrophic future fates or gentle eternal expansion. Equipped with the previous results, Section~\ref{sec:cascade_of_cosmological_rips} presents in a pedagogical manner the possible doomsdays that this DE could predict . In Section~\ref{sec:observational_analysis} we evaluate the observational viability of our models against current cosmological surveys and compare the results with the longstanding concordance model, $\Lambda$CDM.\@ Lastly, we conclude with our final insights in Section~\ref{sec:conclusions}.

% --- The model ---
\section{\label{sec:the_model} The model}

\subsection{Cosmological framework}

For the remainder of this work we consider a cosmological Friedmann-Lemaître-Robertson-Walker spacetime under the assumption of spatial flatness, as suggested by observations from the CMB~\cite{Planck:2019nip,Planck:2018vyg}. More specifically, we consider a universe filled by several species: (baryonic and dark) matter and radiation, and a DE component. Under the assumption that the customary Friedmann and Raychauduri equations hold and the different species only interact gravitationally, the conservation equation for any cosmological fluid $i$ reads%
\footnote{Hereafter we employ reduced Planck units: $c = 8\pi G = 1$.}
\begin{equation}
    \dot{\rho_i} + 3H(\rho_i + p_i) = 0,
    \label{eq:conservation_equation}
\end{equation}
where the overdot stands for differentiation with respect to cosmic time $t$, and $H \equiv \dot{a}/a$ is the Hubble factor. Therefore, matter and radiation have the usual scaling of their energy densities with the scale factor $a$:
\begin{equation}
    \rho_\mathrm{m}(a) =
    3H_0^2 \Omega_\mathrm{m} a^{-3},
    \qquad
    \rho_\mathrm{r}(a) =
    3H_0^2 \Omega_\mathrm{r} a^{-4},
\end{equation}
where $\Omega_i$ denote the current fractional energy densities of the $i$-th species, and $H_0$ corresponds to the Hubble constant. We concentrate on DE, and unless clarity dictates otherwise, $\rho$ and $p$ will represent the energy density and pressure of DE, respectively.

\subsection{Equation of state}

Moving to our proposal, inspired by a general formulation first presented in~\cite{Nojiri:2004pf} and expanded in~\cite{Nojiri:2005sx}, we assume that the pressure and the energy density of DE are related through the following equation:
\begin{equation}
    p =
    -\rho - A \rho^{1-n} \tanh\left( B\rho^n \right),
    \label{eq:pressure}
\end{equation}
where $A$, $B$ and $n$ are real non-zero constants. From this, it immediately follows that the (barotropic) equation-of-state parameter $w \equiv p/\rho$ is given by:
\begin{equation}
    w = -1 - A \rho^{-n} \tanh\left( B\rho^n \right).
    \label{eq:eos}
\end{equation}
Considering that this fluid must act as the agent responsible for the current accelerated expansion of the Universe, the parameters must be compatible with $w < -1/3$. We keep the theoretical analysis general, though we shall come back to this question when contrasting our model with the most up-to-date observations.

Now, employing the conservation equation~\eqref{eq:conservation_equation} for our proposal in Eq.~\eqref{eq:pressure} leads to the following differential equation:
\begin{equation}
    \frac{\mathrm{d} \rho}{\mathrm{d} \ln (a)} =
    3A \rho^{1-n} \tanh\left( B\rho^n \right).
    \label{eq:rho_diff_eq}
\end{equation}
This expression takes advantage of the possibility to resort to derivatives with respect to the scale factor instead of cosmic time, thereby eliminating the explicit dependence on the Hubble parameter $H$. From this form, we see that the parameter $B$ is responsible for controlling the transition from the current value of the energy density to its asymptotic behavior at both early and late times. We initially allowed $B$ to be a free parameter and traced its influence in both the theoretical and observational analyses, and found that leaving $B$ free does not significantly modify any of the main conclusions of this work. However, when $B$ is treated as a free parameter, the analysis of the asymptotic behavior of DE in Sec.~\ref{sec:asymptotic_behavior_of_dark_energy}, as well as the classification of the singularities in Sec.~\ref{sec:cascade_of_cosmological_rips}, depends on the combined sign of the product $AB$, rather than solely on the sign of $A$ that appears explicitly in the main text. For simplicity, and without loss of generality, we therefore opt for fixing $B = {(3H_0^2)}^{-n}$.

Now, in order to maintain a clear visual connection with previous works and at the same time facilitate the comparison with observations, we define the dimensional parameter $\alpha$ so that $A = {(3H_0^2)}^n \alpha$. Setting the current value of the scale factor to $a_0 = 1$ in our equations without loss of generality, and denoting the current value of the fractional energy density as $\Omega_\mathrm{DE}$ so that $\rho(a_0) = 3H_0^2 \Omega_\mathrm{DE}$, the solution reads:
\begin{equation}
    \rho(a) =
    3H_0^2 {\left\{ \ln \!\, \left( \eta a^{3n\alpha} + \sqrt{1 + \eta^2 a^{6n\alpha}} \right) \right\}}^{1/n},
    \label{eq:rho_solution}
\end{equation}
where for convenience we have defined $\eta = \sinh(\Omega_\mathrm{DE}^n)$. Note that the fractional energy density $\Omega_\mathrm{DE}$ is determined by the constraint resulting from evaluating the Friedmann equation at $a = a_0$,
\begin{equation}
    \Omega_\mathrm{m} + \Omega_\mathrm{r} + \Omega_\mathrm{DE} =
    1,
\end{equation}
As current observations suggest that matter and radiation do not saturate the critical energy density ($\Omega_\mathrm{m} + \Omega_\mathrm{r} < 1$), implying $\Omega_\mathrm{DE} > 0$, Eq.~\eqref{eq:rho_solution} reveals that the energy density of DE remains positive throughout cosmic history, i.e., $\rho(a) \geq 0$.

\subsection{General remarks and alternative constructions}

Before moving on, it is important to stress that the proposal in Eq.~\eqref{eq:pressure} is similar to that studied in~\cite{Stefancic:2004kb}, where $p = -\rho - A \rho^{1-n}$ is considered, with the key difference that we are also including the $\rho$-dependent hyperbolic tangent factor. As aforementioned, the energy density of this new formulation of the DE fluid remains positive, yielding a stable evolution that can be confronted with both early- and late-time cosmological observations. On the contrary, the scheme put forward in~\cite{Stefancic:2004kb} suffers from pathological behaviors, which for some combinations of the parameters lead to energy densities that are negative at early times or suddenly become undefined. The plethora of singularities that will arise in our model may be distinct from those previously studied in the literature for the same values of the parameter~\cite{Stefancic:2004kb, Nojiri:2005sx}; nevertheless, the observational basis of our proposal is more robust, as it allows to delineate the cosmological evolution completely without unphysical stages.

We will restrict the discussion to the particular equation-of-state parameter given by Eq.~\eqref{eq:eos}, so as to present the mechanism in a transparent way. One of the main reasons for that specific choice is that analytic expressions of $H(a)$ can be obtained in closed form. Nevertheless, the same phenomenology can be obtained for a wider class of constructions where the corrective term is governed by a sigmoid function of $\rho^n$:
\begin{equation}
    p =
    -\rho - A\rho^{1-n} \sigma(B \rho^n).
\end{equation}
In the latter $\sigma$ is a smooth centered sigmoid function such that $\sigma(0) = 0$, $\sigma^\prime(u) > 0$ and $\sigma(u) \to \sigma_\infty$ as $u \to \infty$. The hyperbolic tangent we will employ in our main proposal is just one of many possible choices, though the conclusions and results of the following sections will remain unaltered given that $\sigma$ is suitably chosen.

% --- Asymptotic behavior of DE ---
\section{\label{sec:asymptotic_behavior_of_dark_energy} \texorpdfstring{Asymptotic behavior of\\ dark energy}{Asymptotic behavior of dark energy}}

Before moving to the classification of future singularities, we first discuss the different asymptotic behaviors of DE predicted by our proposed scenario. As we shall see, the future fate of the Universe, as well as the behavior of DE in its primordial state, are significantly sensitive to the signs of the new parameters $\alpha$ and $n$. In what follows, we analyze in detail the different evolutionary patterns that arise for the different combinations of signs of these two parameters.

\subsection{\label{sec:positive_alpha_and_n} Positive \texorpdfstring{$\alpha$}{α} and \texorpdfstring{$n$}{n}}

From the future singularity perspective, the $\alpha > 0$ and $n > 0$ case is the most interesting one. First, let us stress that this region of the parameter space does not lead to a pathological early-time behavior of DE. In fact, our model predicts:
\begin{equation}
    \lim_{a \to 0} \rho(a) =
    0,
\end{equation}
which guarantees that DE always remains subdominant in the primitive Universe, opening the door to conventional observational tests at high redshifts. Interestingly, in this limit the pressure is related to the energy density as $p \simeq -(1 + \alpha) \rho$, which means that the DE fluid behaves as a phantom component with constant equation-of-state parameter:
\begin{equation}
    \lim_{a \to 0} w(a) =
    -1 - \alpha.
\end{equation}

Hereafter, let us focus on the asymptotic future of the Universe, which we characterize as $a \gg a_0$. Then, the energy density can be approximated as:
\begin{equation}
    \rho(a) \simeq
    3H_0^2 {\big( 3n\alpha \ln(a) \big)}^{1/n} \left\{ 1 + \mathcal{O}\big( 1/\ln(a) \big) \right\}.
\end{equation}
This simplifies considerably the discussion of the future behavior of the Universe, as the only relevant component in the cosmic budget is DE. In this regime, its equation-of-state parameter can also be approximated as:
\begin{equation}
    w(a) \simeq
    -1 - \frac{1}{3n \ln(a)} \left\{ 1 + \mathcal{O}\big( 1/\ln(a) \big) \right\},
\end{equation}
which approaches $-1$ from below as $a \to \infty$. This is a good opportunity to highlight that $w \to -1$ is not a sufficient condition to ensure a de Sitter future, as the energy density for this parameter choice diverges in the asymptotic future. 

The asymptotic behavior of this fluid is equivalent to some of the cases studied in~\cite{Stefancic:2004kb}, and therefore much of the phenomenology discussed there applies here as well. Nevertheless, we pretend to reexamine the possible different rips and future fates from a more general perspective, as well as to provide a more detailed classification of the singularities that arise in this model according to more recent literature.

Now, as the Hubble parameter in the $a \gg a_0$ regime can be approximated as:
\begin{equation}
    H(a) \simeq
    \sqrt{3} H_0 {\left\{ \ln \!\, \left( \eta a^\gamma + \sqrt{1 + \eta^2 a^{2\gamma}} \right) \right\}}^{1/(2n)},
\end{equation}
we can obtain further dynamical information by computing its time derivatives. Recalling that $\mathrm{d} / \mathrm{d} t = aH \, \mathrm{d} / \mathrm{d} a$, we can, for instance, obtain the first time derivative:
\begin{equation}
    \dot{H}(a) \simeq
    \frac{{(3n\alpha)}^{1/n}}{2n} H_0^2 {(\ln a)}^{(1 - n)/n}.
\end{equation}
Higher order time derivatives can be computed in a similar fashion. More specifically, the $k$-th time derivative can be expressed by the following general formula:
\begin{equation}
    \frac{\mathrm{d}^k H}{\mathrm{d} t^k} \simeq
    H_0^{k+1} {(3n\alpha)}^{\frac{k+1}{2n}} {(\ln a)}^{\frac{k+1}{2n}-k} \prod_{j=0}^{k-1} \left( \frac{j+1}{2n} - j \right).
    \label{eq:time_derivative_H}
\end{equation}
Eq.~\eqref{eq:time_derivative_H} is particularly interesting, as it allows for a general characterization of the future singularity structure of the model depending on the values of $n$. First of all, note that all $k \geq m$ derivatives vanish regardless of the value of the scale factor when
\begin{equation}
    n = \frac{m+1}{2m},
    \qquad
    m \in \mathbb{Z}^+/\{1\}.
    \label{eq:n_values_no_singularity}
\end{equation}
It is worth remarking that such vanishing time derivatives arise only when the parameter lies in the interval $1/2 < n < 1$. Now, let us assume that $n$ does not take any of the previous values. Then, in view of Eq.~\eqref{eq:time_derivative_H}, the $k$-th time derivative of the Hubble parameter, $H^{(k)}$, can have three different asymptotic behaviors depending on the value of the exponent $n$: i) $H^{(k)}$ blows up to infinity when $n < (k+1)/(2k)$; ii) $H^{(k)}$ tends to a constant value when $n = (k+1)/(2k)$; and iii) $H^{(k)}$ vanishes when $n > (k+1)/(2k)$.

The latter is a remarkable result, as it shows that the model can lead to a rich variety of future scenarios. More specifically, if $n < 1/2$, then all time derivatives of the Hubble parameter diverge as $a \to \infty$; if $1/2 < n < 1$, then there exists a finite number of time derivatives that diverge. Lastly, if $n > 1$, then all time derivatives tend to zero as $a \to \infty$, yielding the softest type of singularity.

\subsection{Positive \texorpdfstring{$\alpha$}{α} and negative \texorpdfstring{$n$}{n}}

This case also yields interesting future scenarios. Before moving to the late-time regime, we first discuss the early-time behavior ($a \ll a_0$). In this limit, the energy density can be approximated as:
\begin{equation}
    \rho(a) \simeq
    3H_0^2 /{\big(\hspace{-3pt}-\hspace{-2pt} 3\vert n \vert\alpha \ln(a) \big)}^{-1/\vert n \vert},
\end{equation}
and therefore we notice that $\rho(a) \to 0$ as $a \to 0$. Unlike the positive $n$ scenario, the equation-of-state parameter asymptotically approaches to that of a cosmological constant from below the phantom divide line
:
\begin{equation}
    w(a) \simeq
    -1 + \frac{1}{3\vert n \vert \ln(a)}.
\end{equation}

Moving to the asymptotic future ($a \gg a_0$), the energy density can be approximated as:
\begin{equation}
    \rho(a) \simeq
    3H_0^2 \eta^{-1/\vert n \vert} a^{3\alpha},
\end{equation}
showing that it diverges as $a \to \infty$. This evolution is identical to a phantom fluid with constant equation of state $w(a) \simeq -1 - \alpha$, and as we shall identify later, it leads to the so well-known Big Rip singularity.

\subsection{Non-singular future fates (\texorpdfstring{$\alpha < 0$}{α<0})}

The remaining two sign combinations, both corresponding to $\alpha < 0$, lead to inverted early- and late-time behaviors compared to the previous cases we have discussed. This means that the energy density diverges at early times but dilutes asymptotically. This is expected since, as suggested by Eq.~\eqref{eq:eos} it turns out that $w$ always remains above the phantom divide line. For completeness, we briefly review these scenarios.

\subsubsection{Negative \texorpdfstring{$\alpha$}{α} and positive \texorpdfstring{$n$}{n}}

In the early-universe regime, the DE blows up logarithmically:
\begin{equation}
    \rho(a) \simeq
    3H_0^2 {\big(\hspace{-3pt}-\hspace{-2pt} 3n\vert \alpha \vert \ln(a) \big)}^{1/n}.
\end{equation}
This logarithmic divergence is much milder than the polynomial scalings of matter and radiation (which behave as $a^{-3}$ and $a^{-4}$, respectively), so for any fixed $n > 0$ and reasonable $\vert \alpha \vert$ it does not spoil the standard matter-radiation dominated era. On the other hand, at late times we find:
\begin{equation}
    \rho(a) \simeq
    3H_0^2 \eta^{1/n} a^{-3\vert \alpha \vert},
\end{equation}
which scales as a quintessence fluid with constant equation-of-state parameter $w \simeq -1 + \vert \alpha \vert$. For reasonable values of $\alpha$ so that $w < 0$, this fluid dilutes with cosmic expansion though at a much lower rate than matter and radiation. Therefore, the cosmic expansion continues forever with $H \to 0$.

\subsubsection{Negative \texorpdfstring{$\alpha$}{α} and \texorpdfstring{$n$}{n}}

The conclusions of the previous scenario are also valid in this case. In fact, when $a \ll a_0$, the scaling of DE is of the form
\begin{equation}
    \rho(a) \simeq
    3H_0^2 \eta^{-1/\vert n \vert} a^{-3\vert \alpha \vert},
\end{equation}
thus diverging as a quintessence fluid with $w \simeq -1 + \vert \alpha \vert$. For a reasonable early-time description of the Universe (dominated both by matter and radiation), one requires $\vert \alpha \vert < 1$. In view of to the late-time behavior of DE, we find:
\begin{equation}
    \rho(a) \simeq
    3H_0^2 / {\big( 3n\alpha \ln(a) \big)}^{1/\vert n \vert},
\end{equation}
thus diluting logarithmically with the expansion. In this case we also find that DE is the dominant component in the asymptotic future, though the expansion rate continues to slow down until $H \to 0$ when $a \to \infty$.

% --- Cascade of cosmological rips ---
\section{\label{sec:cascade_of_cosmological_rips} Cascade of cosmological rips}

In the previous section we have analyzed the different asymptotic regimes of the proposed DE, and we have found that singular behaviors associated with a divergence in the energy density arise in the future when $\alpha > 0$. Now, we shall discuss the nature of these singularities by studying the time dependence of the scale factor depending on the values adopted by the exponent $n$. For the remainder of the discussion, we neglect matter and radiation components, as they are subdominant for $a \gg a_0$ compared with DE. It should be noted that our construction of the energy density, Eq.~\eqref{eq:rho_solution}, does not allow for a closed-form expression of the scale factor $a(t)$, so all equations below should be interpreted as describing the approximate behavior near the singularity. The $\alpha < 0$ region is excluded from the following analysis as the Universe undergoes an eternal expansion with an asymptotically vanishing Hubble factor.

\subsection{Big Rip (\texorpdfstring{$n \leq 0$}{n≤0})}

We have seen that these scenarios predict a DE fluid behaving as a phantom component with constant equation of state. The case $n = 0$ is also included, since from Eq.~\eqref{eq:eos} the fluid likewise exhibits a phantom characterization. This implies that the Hubble function scales as
\begin{equation}
    {H(a)}^2 \sim
    3H_0^2 \eta^{-1/\vert n \vert} a^{3\alpha}.
\end{equation}
This simple power-law dependence allows to easily obtain the scaling of the scale factor $a$ in terms of the cosmic time $t$. The above equation can be straightforwardly integrated, yielding
\begin{equation}
    a(t) \sim
    {\left( \frac{2}{3\alpha} H_0 (t_\mathrm{rip} - t) \right)}^{-2/(3\alpha)}.
\end{equation}
Thus, the scale factor diverges at a finite future time $t = t_\mathrm{rip}$. This corresponds to the well-known Big Rip~\cite{Caldwell:2003vq, McInnes:2001zw}, or Type I singularity~\cite{Nojiri:2005sx}, which has driven much of the interest in the study of cosmological singularities.

\subsection{Grand Rip (\texorpdfstring{$0 < n < 1/2$}{0<n<1/2})}

Recalling our earlier approximate expression for the energy density in the $n > 0$ regime, the Hubble parameter can be expressed as
\begin{equation}
    H(a) \sim
    H_0 {\big( 3n\alpha \ln(a) \big)}^{1/(2n)},
    \label{eq:H_positive_n}
\end{equation}
which can be integrated to obtain
\begin{equation}
    \ln a(t) \sim
    {\left[ \left( \frac{1}{2n} - 1\right) {(3n\alpha)}^{1/(2n)} H_0(t_\mathrm{rip} - t) \right]}^{2n/(2n-1)}
    \label{eq:scale_factor_time}
\end{equation}
for $n \neq 1/2$. Since the exponent in the above time relation is negative, the scale factor $a(t)$, the Hubble parameter $H(t)$ and all its time derivatives diverge at a finite cosmic time $t = t_\mathrm{rip}$. Interestingly, the equation-of-state parameter for these values of $n$ evolves as:
\begin{equation}
    1 + w(t) \propto
    {(t_\mathrm{rip} - t)}^{\frac{2n}{1 - 2n}},
\end{equation}
approaching $w \to -1$ in the singularity limit. We identify this future fate with a Grand Rip~\cite{Fernandez-Jambrina:2014sga} --- a tempered version of the Big Rip singularity in which $w$ evolves smoothly toward that of a cosmological constant. According to the classification of~\cite{Nojiri:2005sx}, this singularity also falls into the Type~I category.

\subsection{Mild Rip (\texorpdfstring{$n = 1/2$}{n=1/2})}

This is a borderline case, as the solution for the scale factor differs from that in Eq.~\eqref{eq:scale_factor_time}. Solving the differential equation~\eqref{eq:H_positive_n} shows that the scale factor scales as a double exponential of the cosmic time:
\begin{equation}
    a(t) \sim
    \exp\left\{ \exp\left( \frac{3}{2\alpha} H_0(t - t_\ast) \right) \right\}.
\end{equation}
Here $t_\ast$ denotes an arbitrary time scale for which the description of the Universe solely in terms of DE becomes appropriate. Now, even though the scale factor grows extremely fast, the singularity is postponed to infinite cosmological time. Similarly, the Hubble factor and all its time derivatives diverge as $t \to \infty$. Reference~\cite{Frampton:2011sp} included this borderline case into the Little Rip category (the next case down the ladder) and studied the boundary between finite- and infinite-time singularities in detail. This scenario can be viewed as a super-violent Little Rip, though not as abrupt as the Big/Grand Rip. Therefore, based on  the general characterization of the time derivatives of the Hubble parameter, Eq.~\eqref{eq:time_derivative_H}, which all diverge, we dub this singularity as Mild Rip.

\subsection{Little Rip (\texorpdfstring{$1/2 < n < 1$}{1/2<n<1})}

In this range of values of $n$, the scale factor $a(t)$ can be expressed as
\begin{equation}
    \ln a(t) \sim
    {\left[ \left( 1 - \frac{1}{2n} \right) H_0 {(3n\alpha)}^{1/(2n)} (t - t_\ast) \right]}^{2n/(2n - 1)},
    \label{eq:scale_factor_little_rip}
\end{equation}
and therefore the Hubble parameter reads
\begin{equation}
    H(t) \sim
    H_0 {\left( t - t_\ast \right)}^{1/(2n - 1)}.
    \label{eq:hubble_little_rip}
\end{equation}
The latter approximations are valid as long as $t \gg t_\ast$. As the exponent in the scale factor is positive, in this case both the scale factor $a(t)$ and the Hubble parameter $H(t)$ diverge at infinite cosmic time. However, from Eq.~\eqref{eq:time_derivative_H} we see that the first $\lfloor {(2n - 1)}^{-1} \rfloor$ time derivatives of the Hubble parameter also diverge as $t \to \infty$ ($\lfloor \dots \hspace{-0.4pt}\rfloor$ stands for the floor function), whereas higher order derivatives vanish. Bound structures are inevitably disrupted in a finite future time. The location of this disruption, however, depends on the model parameter 
$n$: larger values of 
$n$ shift the rip time further into the future. We identify this softer singularity with a Little Rip~\cite{Frampton:2011sp}; many physical models reproduce this scenario, see e.g.~\cite{Frampton:2011rh, Brevik:2011mm}.

Here we mention that a more precise classification of this rip scenario could be developed in correspondance to the number of non-zero derivatives of the Hubble parameter. This has been the case, for example, when proposing the Little Sibling of the Big Rip (our next study case), which was previously classified into the Little Rip category without deepening further in the $H^{(k)}$. Nevertheless, since there are an infinite number of subcases to discuss, we consider appropriate to preserve the standard labeling from the literature.

\subsection{Little Sibling of the Big Rip \texorpdfstring{$\boldsymbol{(n = 1)}$}{(n=1)}}

For this case, one finds that the scale factor evolves as
\begin{equation}
    \ln a(t) \sim
    \frac{3}{4}\alpha H_0^2 {(t - t_\ast)}^2,
\end{equation}
and that the Hubble factor simply grows linearly with time,
\begin{equation}
    H(t) \sim
    \frac{3}{2}\alpha H_0^2 (t - t_\ast).
\end{equation}
It then immediately follows that the first derivative of the Hubble parameter is constant, $\dot{H} \sim (3/2)\alpha H_0^2$, whereas higher order derivatives vanish. This singularity is softer than a Little Rip, and in the literature it is known as the Little Sibling of the Big Rip~\cite{Bouhmadi-Lopez:2014cca}.

\begin{table*}[t]
    \centering
    \setlength{\tabcolsep}{1.25em}
    \begin{tabular}{ccccc}
        \hline\hline\rule{0pt}{1.em}%
        Scenario & $n$ & $t_\mathrm{rip}$ & Asymptotic $H(t)$ & Asymptotic $H^{(k)}(t)$ \\ \hline\rule{0pt}{1.2em}%
        Big Rip & $(-\infty, 0]$ & Finite & $t^{-2/(3\alpha)}$ & $\infty$ \\[0.7em]
        Grand Rip & $(0, 1/2)$ & Finite & $\exp\big( t^{-\frac{1}{1-2n}} \big)$ & $\infty$ \\[0.7em]
        Mild Rip & $1/2$ & $\infty$ & $e^t$ & $\infty$ \\[0.7em]
        Little Rip & $(1/2, 1)$ & $\infty$ & $\sqrt[2n-1]{t}$ & $\infty$ for $k < \lfloor (2n - 1)^{-1} \rfloor$ \\[0.7em]
        Little Sibling of the Big Rip & $1$ & $\infty$ & $t$ & $(3\alpha/2) H_0$ \\[0.7em]
        Dollhouse Rip & $(1, \infty)$ & $\infty$ & $\sqrt[2n-1]{t}$ & $0$ \\%
        \hline\hline
    \end{tabular}
    \caption{Summary of the different rip scenarios generated by our equation-of-state parameter~\eqref{eq:eos}. The second column indicates the range of values of the parameter $n$ leading to each scenario. The third column states whether the singularity occurs at finite or infinite cosmic time. The fourth column describes the asymptotic behavior of the Hubble parameter, whereas the fifth column shows the asymptotic behavior of its time derivatives. For the precise characterization of the asymptotic $H^{(k)}$ in the Little Rip scenario, see the detailed discussion in Section~\ref{sec:positive_alpha_and_n}.}%
    \label{tab:rip_scenarios}
\end{table*}

\subsection{Dollhouse Rip \texorpdfstring{$\boldsymbol{(n > 1)}$}{(n>1)}}

Finally, for $n > 1$ the scale factor $a(t)$ and the Hubble factor $H(t)$ evolve as dictated by Eqs.~\eqref{eq:scale_factor_little_rip} and~\eqref{eq:hubble_little_rip}, respectively. Even though both quantities diverge as $t \to \infty$, the Hubble parameter in this case grows sublinearly with time and consequently all its time derivatives vanish in this limit. The expansion is always accelerated, but it happens in the gentlest possible way among the phantom scenarios reproduced by our model, with vanishing $H^{(k)}$. In view of  the latter, we dub this the Dollhouse Rip. Even though in the literature this rip was categorized as a Little Sibling of the Big Rip (see e.g.~\cite{BorislavovVasilev:2021srn}), we prefer to reserve this label for the $\dot{H} = \mathrm{constant}$ case and classify the $H \to \infty$, $H^{(k)} \to 0$ ($k \in \mathbb{Z}^+ / \{1\}$) as a distinct rip.

\subsection{General remarks}

Even though all these cosmological frameworks have been thoroughly studied in the literature from a theoretical standpoint, our equation-of-state parameter generates a unified family of ``Rip Cosmologies'' while at the same time avoids early-time pathological behaviors. This makes this DE framework be worth confronting with astronomical probes to fill existing gaps in  systematic assessments based on precision data. A summary of the different cosmological rips arising from our model is provided in Table~\ref{tab:rip_scenarios}.

% --- Observational analysis ---
\section{\label{sec:observational_analysis}Observational analysis}

\subsection{Bayesian inference and datasets}

In order to test the viability of our proposed models, we perform a Bayesian analysis using the latest cosmological data. More specifically, we use the following four precision datasets:
\begin{itemize}
    \item \emph{Baryon Acoustic Oscillations} (\textbf{BAO}): We make use of the latest measurements from the Data Release 2 of the DESI~\cite{DESI:2025zgx}. We include a measurement of the angle-averaged distance $D_\mathrm{V}/r_\mathrm{d}$ at an effective redshift $z_\mathrm{eff} \!=\! 0.295$, and six measurements of the transverse and Hubble distances $D_\mathrm{M}/r_\mathrm{d}$ and $D_\mathrm{H}/r_\mathrm{d}$ at effective redshifts $0.51 \leq z_\mathrm{eff} \leq 2.33$; their corresponding covariance matrix is computed using the values in~\cite{DESI:2025zgx}. The comoving sound horizon at the drag epoch $r_\mathrm{d}$ is obtained from the photon-baryon fluid dynamics using the fitting formula for the redshift at the drag epoch $z_\mathrm{d}$ from~\cite{Aizpuru:2021vhd}.
    
    \item \emph{Cosmic Chronometers} (\textbf{CC}): We consider the compilation of measurements from~\cite{Moresco:2022phi} of the Hubble parameter in the redshift interval $0 < z < 1.965$. These measurements are provided by the differential age method applied to massive and passively evolving galaxies~\cite{Jimenez:2001gg, Moresco:2010wh, Moresco:2018xdr, Moresco:2020fbm, Moresco:2022phi} that provide model independent measurements of $H(z)$ under the assumption of a FLRW spacetime~\cite{Moresco:2012by, Moresco:2015cya, Moresco:2012jh, Moresco:2016nqq, Moresco:2017hwt, Jimenez:2019onw, Jiao:2022aep}. We account for the correlations among measurements as exemplified in~\cite{Moresco:2020fbm}.

    \item \emph{Cosmic Microwave Background} (\textbf{CMB}): As the proposal does not significantly affect the early-universe dynamics, we are allowed to consider the compressed likelihood of the CMB from~\cite{Bansal:2025ipo}.\@ It consists of measurements of the shift parameters $R$ and $l_a$, introduced in~\cite{Wang:2007mza}, and the reduced fractional baryon density $\Omega_\mathrm{b} h^2$, where $h = H_0/(100\, \mathrm{km}\, \mathrm{s}^{-1}\, \mathrm{Mpc}^{-1})$. To compute the shift parameters, we use the fitting formula for the recombination redshift $z_\ast$ from~\cite{Aizpuru:2021vhd}.
    
    \item \emph{Type Ia Supernovae} (\textbf{SN}): We use the latest Pantheon+ compilation~\cite{Scolnic:2021amr, Peterson:2021hel, Carr:2021lcj, Brout:2022vxf} consisting of 1701 light curves from 1550 distinct SN, which we select to lie in the redshift range $0.01 < z < 2.26$ to reduce the impact of peculiar velocities~\cite{Peterson:2021hel} of very low redshift supernovae. We decide not to include \emph{SH0ES}'s calibrated Cepheids as these are inconsistent with the CMB~\cite{Brout:2022vxf}, ultimately requiring marginalization over the absolute fiducial magnitude~\cite{Conley_2010}.
\end{itemize}

\begin{table}[t]
    \centering
    \setlength{\tabcolsep}{1em}
    \begin{tabular}{ll}
        \hline\hline\rule{0pt}{.75em}%
        Parameter & Prior (flat) \\ \hline\rule{0pt}{1.em}%
        $\Omega_\mathrm{m}$ & $\mathcal{U}[0.2, 0.8]$ \\[0.3em]
        $H_0 $ & $\mathcal{U}[20, 100]$ \\[0.3em]
        $\Omega_\mathrm{b} h^2$ & $\mathcal{U}[0.005, 0.1]$ \\[0.3em]
        $\alpha$ & No prior \\ \hline\hline 
    \end{tabular}
    \caption{Prior distributions for the sampled parameters; $\mathcal{U}[x, y]$ refers to a uniform distribution over the interval $[x, y]$.}%
    \label{tab:priors}
\end{table}

We combine all previous datasets and perform Monte Carlo Markov Chain (MCMC) analyses, sampling over the cosmological parameters $(\Omega_\mathrm{m}, H_0, \Omega_\mathrm{b}, \alpha)$ under the flat priors summarized in Table~\ref{tab:priors}. We then obtain the resulting ``best-fit'' (more specifically, the median) values for different choices of the exponent $n$ characterizing the different rip scenarios. Allowing the exponent $n$ to vary freely leads to largely unconstrained results in the $(\alpha, n)$ parameter space; for this reason, we fix it to representative values of future rips summarized in Table~\ref{tab:rip_scenarios}. In addition, despite $\alpha < 0$ not leading to future singularities, we decide not to impose any prior on $\alpha$ to allow for the possibility of both phantom or quintessence DE.

To asses the statistical significance of our models, we consider $\Lambda$CDM as the benchmark model. We do this by computing the Bayes' factor, defined as the ratio between the Bayesian evidences of the $\Lambda$CDM and the model with exponent $n$: 
\begin{equation}
    Z_{\Lambda\mathrm{CDM}}^n \equiv Z_n/Z_{\Lambda\mathrm{CDM}}.
\end{equation}
Positive values indicate evidence in favor of our model. The evidences are computed using the Nested Sampling algorithm described in~\cite{Mukherjee:2005wg} and interpreted according to Jeffrey's scale~\cite{Jeffreys61}: values of $0 \leq \vert \ln Z_n^{\Lambda\mathrm{CDM}} \vert < 1$ indicate weak evidence, $1 \leq \vert \ln Z_n^{\Lambda\mathrm{CDM}} \vert < 2.5$ substantial evidence, $2.5 \leq \vert \ln Z_n^{\Lambda\mathrm{CDM}} \vert < 5$ strong evidence, and $\vert \ln Z_n^{\Lambda\mathrm{CDM}} \vert \geq 5$ decisive evidence.

\subsection{Results}

\begin{table*}[ht]
    \centering
    \setlength{\tabcolsep}{.55em}
    \begin{tabular}{cccccccc}
        \hline\hline\rule{0pt}{1em}%
        Model & $\Omega_\mathrm{m}$ & $H_0$ & $\Omega_\mathrm{b}$ & $\alpha$ & $w_0$ &$\chi_{\min}^2$ & $\ln Z_{\Lambda\mathrm{CDM}}^n$ \\ \hline\rule{0pt}{1.25em}%
        $\Lambda$CDM	 & 	$0.3020_{-0.0036}^{+0.0037}$	 & 	$68.27_{-0.28}^{+0.28}$	 & 	$0.04836_{-0.00032}^{+0.00032}$	 & --------	 & $-1$ & 	$1448.07$	 & 	--------	 \\[1.4em]
        $n = -1/2$ (BR)	 & 	$0.3043_{-0.0050}^{+0.0050}$	 & 	$67.94_{-0.56}^{+0.56}$	 & 	$0.04890_{-0.00086}^{+0.00087}$	 & 	$-0.022_{-0.032}^{+0.033}$ & $-0.9985_{-0.0023}^{+0.0023}$	 & 	$1447.63$	 & 	$-0.176_{-0.029}^{+0.029}$	 \\[1.4em]
        $n = 1/3$ (GR)	 & 	$0.3043_{-0.0050}^{+0.0052}$	 & 	$67.92_{-0.57}^{+0.58}$	 & 	$0.04892_{-0.00087}^{+0.00087}$	 & 	$-0.020_{-0.028}^{+0.030}$ & $-0.9944_{-0.0082}^{+0.0079}$	 & 	$1447.63$	 & 	$-0.133_{-0.028}^{+0.029}$	 \\[1.4em]
        $n = 1/2$ (MR)	 & 	$0.3044_{-0.0051}^{+0.0050}$	 & 	$67.92_{-0.56}^{+0.59}$	 & 	$0.04892_{-0.00087}^{+0.00088}$	 & 	$-0.020_{-0.028}^{+0.029}$	& $-0.9971_{-0.0042}^{+0.0040}$ & 	$1447.61$	 & 	$-0.149_{-0.028}^{+0.028}$	 \\[1.4em]
        $n = 2/3$ (LR)	 & 	$0.3044_{-0.0050}^{+0.0051}$	 & 	$67.92_{-0.58}^{+0.58}$	 & 	$0.04892_{-0.00087}^{+0.00089}$	 & 	$-0.020_{-0.028}^{+0.028}$ & $-0.9985_{-0.0022}^{+0.0022}$	 & 	$1447.62$	 & 	$-0.145_{-0.026}^{+0.028}$	 \\[1.4em]
        $n = 1$	(LSBR) & 	$0.3046_{-0.0052}^{+0.0051}$	 & 	$67.90_{-0.58}^{+0.59}$	 & 	$0.04895_{-0.00089}^{+0.00090}$	 & 	$-0.019_{-0.027}^{+0.028}$ & $-0.99959_{-0.00058}^{+0.00059}$	 & 	$1447.60$	 & 	$-0.159_{-0.029}^{+0.031}$	 \\[1.4em]
        $n = 4/3$ (DR)	 & 	$0.3044_{-0.0050}^{+0.0051}$	 & 	$67.92_{-0.57}^{+0.58}$	 & 	$0.04894_{-0.00086}^{+0.00090}$	 & 	$-0.019_{-0.026}^{+0.026}$ & $-0.99989_{-0.00015}^{+0.00016}$	 & 	$1447.61$	 & 	$-0.125_{-0.030}^{+0.031}$ \\[0.3em] \hline\hline
    \end{tabular}
    \caption{Best fit values and uncertainties for the concordance $\Lambda$CDM model and one representative per rip scenario presented in Section~\ref{sec:cascade_of_cosmological_rips}. Only the values for the combination CMB+BAO+CC+SN are reported. We also include the current value of the equation of state $w_0$, the minimum $\chi^2$ for each run, $\chi_{\min}^2$, and the logarithm of the Bayes' factors $\ln Z_n^{\Lambda\mathrm{CDM}}$. The initials of the first column are as follows: Big Rip (BR); Grand Rip (GR); Mild Rip (MR); Little Rip (LR); Little Sibling of the Big Rip (LSBR); and Dollhouse Rip (DR).}%
    \label{tab:results}
\end{table*}  

Table~\ref{tab:results} contains the main observational results of this model. In order to contemplate all possible future fates of the Universe, we have selected one value of the exponent $n$ per rip scenario analyzed in Section~\ref{sec:cascade_of_cosmological_rips}. The best fit values for all models are compatible with the concordance $\Lambda$CDM model within $1\sigma$. We observe higher values of $\Omega_\mathrm{m}$ compared with the best fit for $\Lambda$CDM, and lower values of the Hubble parameter $H_0$. The latter, due to the Gaussian multivariate prior from the CMB over the shift parameters and $\Omega_\mathrm{b} h^2$, results in slightly larger values of the fractional baryon density $\Omega_\mathrm{b}$. The statistical significance, encoded in the logarithm of the Bayes' factor, indicates weak evidence in favor of $\Lambda$CDM according to Jeffreys' scale.

Further analysis of Table~\ref{tab:results} shows that all models, classified by the exponent $n$, present equal best-fit values for all sampled parameters, including the new parameter $\alpha$. Given the compatibility of $\alpha$ with the value of zero, and based on the expression of the energy density of DE~\eqref{eq:rho_solution}, it turns out that the latter scales as
\begin{equation}
    \frac{\rho(a)}{3H_0^2} \simeq
    \Omega_\mathrm{DE} - \mathcal{O}\big( \alpha \log (1+z) \big).
    \label{eq:rho_approximate}
\end{equation}
The proportionality constants in the correction terms are functions of powers of $\Omega_\mathrm{DE}$ which are largely $n$-independent when $\vert \alpha \vert \ll 1$. On the other hand, the time-evolution in terms of the redshift is very slow, and therefore late-time surveys $(z \lesssim 3)$ are almost insensitive to the evolution of DE.\@ Consequently, the numerical analysis is unable to distinguish between the potential cosmological scenarios. As mentioned earlier in the text, we have conducted an analysis in which we permitted $n$ to vary, though the constraints are poor (in a statistical sense), indicating that the MCMC method does not converge to any particular value of $n$.

\begin{figure}[t]
    \centering
    \includegraphics{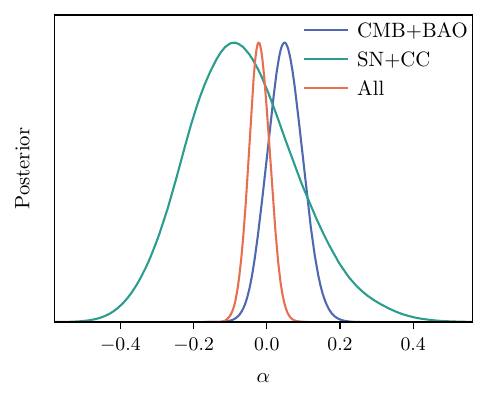}
    \vspace{-1em}
    \caption{One-dimensional marginalized and normalized posterior distributions of $\alpha$ for $n = 1/3$. As the results for $\alpha$ are almost $n$-independent, similar distributions follow for other rip scenarios. Datasets are combined  into early- (CMB+BAO) or late-time (SN+CC) observables.}%
    \label{fig:alpha_distribution}
\end{figure}

In order to estimate the impact of distinct datasets in the measured value of $\alpha$, we performed two additional MCMCs distinguishing data that involve early-time physics, that is, the CMB and BAO, and late-time physics, corresponding to SN and CC.\@ The same pattern from the combined analysis emerges: the results for the sampled parameters are largely $n$-independent. Therefore, we opt for selecting a particular representant among these, such as $n = 1/3$, and illustrate the one-dimensional marginalized posterior distribution for the parameter $\alpha$ in Fig.~\ref{fig:alpha_distribution}. We observe that early-time data seem to favor slightly positive values of $\alpha$, whereas negative values are preferred by late-time surveys although with much larger uncertainties. As aforementioned, the slow (logarithmic) evolution of DE makes SN and CC largely insensitive to its time-dependence (and thus to the $\alpha$ dependence); this is surprising, as DE is commonly more strongly constrained at low-redshift.

Table~\ref{tab:results} also reports the median and asymmetric errors of the current value of the equation-of-state parameter, $w(a_0) \equiv w_0$. Consistently with the results of $\alpha$, the $w_0$ indicate that DE behaves as a quintessence fluid, although with an equation of state compatible with no evolution --- a cosmological constant --- within $1\sigma$. As aforementioned, DE evolves logarithmically with the scale factor across all cosmic history, and since by construction we are only able to explore the $a \leq 1$ region, the time variation of DE is unnoticeable by surveys. Moreover, recent data seem to require a rapidly evolving DE --- as suggested both by parametric equations of state and non-parametric reconstructions of the energy density ---, but our proposal remains close to $\Lambda$CDM.\@ This indicates that DE models which pretend to predict future cosmological rips and at the same time be statistically supported by observations over $\Lambda$CDM would need to allow for more complex dynamics in the recent past.

% === Conclusions ===
\section{\label{sec:conclusions} Conclusions}

The concern about what lies ahead in time
has been a major intellectual driver of cosmology. The discovery of the enigmatic component driving the current accelerated expansion of the Universe opened the door to an array of possible cosmic destinies, stimulating theoretical interest in the ultimate fate of the Universe. In parallel, the steadily increasing precision of cosmological datasets is bringing us closer to determining the precise dynamics of the Universe from its earliest moments to the present day---and perhaps even to probing its future evolution. 

In order to do so, a common and very effective practice has been the exploration of DE parameterizations with minimal but clear theoretical motivations. Given the recent observational trend favoring evolving DE, confronting rip models with up-to-date surveys becomes relevant in order to shed some light on this topic.
Nevertheless, earlier proposals concerned with future cosmological rips commonly ignored observational cosmology, yielding descriptions of DE which were ill-defined at early times.

The aim of our work was to provide a well-behaved evolution of DE throughout all cosmic history, with a positive energy density at all times, which at the same time could provide a cascade of cosmological rips of theoretical interest. Our study case, encoded in Eq.~\eqref{eq:eos}, stems from the proposal in~\cite{Nojiri:2004pf,Nojiri:2005sx, Stefancic:2004kb} with the addition of a sigmoid function which smooths the evolution of DE at all epochs. For some sign combinations of the new parameters $\alpha$ (responsible of the quintessence or phantom characterization of DE) and $n$ (responsible for the classification of singularities), our proposal was able to bring a cascade of possible cosmological rips. Summarized in Table~\ref{tab:rip_scenarios}, those are the well-known Big Rip, the Grand Rip, the Mild Rip and Little Rip (the former commonly labeled also as Little Rip), the Little Sibling of the Big Rip, and the Dollhouse Rip (usually categorized as a Little Sibling of the Big Rip).

Given the consistent time-evolution of our proposal, we were able to observationally test our model with the most up-to-date early- (CMB and BAO) and late-time (SN and CC) datasets. A thorough statistical analysis employing MCMC pipelines yield best-fit results compatible with those of the concordance model ($\Lambda$CDM) within one standard deviation. Further statistical tests based on the use of the Bayes factor (as known as Bayesian evidence ratio) indicate weak evidence~\cite{Jeffreys61} favoring $\Lambda$CDM.\@ The reason behind these findings can be attributed to the slow logarithmic evolution of the energy density of DE with the scale factor $a$. As by definition observations can only prove the Universe for $a \in (0,1)$, the model is unable to exhibit the strong time-evolution suggested by recent analyses from the DESI collaboration~\cite{DESI:2025zgx}, yielding a quasi-constant equation of state. For the same reason, all possible characterizations of the future fate, encoded in the exponent $n$, seem to be equally favored.

In conclusion, the most popular characterizations of a plethora of violent future fates lack more complex dynamics in the local universe ($z \lesssim 2.5$) to be compliant with current observational surveys. Even though our proposal yields a perfectly valid evolution of DE at all times, abrupt changes in the equation-of-state parameter (presumably including a crossing of the phantom divide line $w = -1$) should be contemplated in order for these rip models to be statistically favored over $\Lambda$CDM.\@ Yet, the background geometrical tests of the evolution are insensitive to whether the Universe has an abrupt final fate or fits in  the $\Lambda$CDM paradigm.

% --- Acknowledgments ---
\begin{acknowledgments}
MA acknowledges support from the Basque Government Grant No.~PRE\_2024\_1\_0229. MA and RL are supported by the Basque Government Grant IT1628-22, and by Grant PID2021-123226NB-I00 (funded by MCIN/AEI/10.13039/501100011033, by ``ERDF A way of making Europe''). This article is based upon work from COST Actions CosmoVerse CA21136 and CaLISTA CA21109, supported by COST (European Cooperation in Science and Technology). 
\end{acknowledgments}

% --- Bibliography ---
% For titles in APS, download the custom .bst from the following URL:
% https://gist.github.com/goerz/010379874fbff3cc568fbb9fdf63888c
% To activate arXiv links, search for "FUNCTION {format.eprint.controlled}", 
% uncomment the braces below and comment the { "" } before "FUNCTION {format.eprint}"
\bibliographystyle{./apsrev4-2-titles}
\bibliography{./Bibliography}

\end{document}